# Throughput Analysis of CSMA Wireless Networks with Finite Offered-load

Caihong Kai † and Shengli Zhang *
†School of Computer and Information, Hefei University of Technology, Hefei, China
*Department of Communication Engineering, Shenzhen University, China
Email:chkai@hfut.edu.cn, zsl@szu.edu.cn

*Abstract*—This paper proposes an approximate method, equivalent access intensity (EAI), for the throughput analysis of CSMA wireless networks in which links have finite offered-load and their MAC-layer transmit buffers may be empty from time to time. Different from prior works that mainly considered the saturated network, we take into account in our analysis the impacts of empty transmit buffers on the interactions and dependencies among links in the network that is more common in practice. It is known that the empty transmit buffer incurs extra waiting time for a link to compete for the channel airtime usage, since when it has no packet waiting for transmission, the link will not perform channel competition. The basic idea behind EAI is that this extra waiting time can be mapped to an equivalent "longer" backoff countdown time for the unsaturated link, yielding a lower link access intensity that is defined as the mean packet transmission time divided by the mean backoff countdown time. That is, we can compute the "equivalent access intensity" of an unsaturated link to incorporate the effects of the empty transmit buffer on its behavior of channel competition. Then, prior *saturated* ideal CSMA network (ICN) model can be adopted for link throughput computation. Specifically, we propose an iterative algorithm, "Compute-and-Compare", to identify which links are unsaturated under current offered-load and protocol settings, compute their "equivalent access intensities" and calculate link throughputs. Simulation shows that our algorithm has high accuracy under various offered-load and protocol settings. We believe the ability to identify unsaturated links and compute links throughputs as established in this paper will serve an important first step toward the design and optimization of general CSMA wireless networks with offered-load control.

*Index Terms*—Throughput computation, Unsaturated, Offered-load, CSMA, IEEE 802.11.

## I. Introduction

With the widespread deployment of IEEE 802.11 networks, it is common today to find multiple wireless LANs co-located in the neighborhood of each other. The multiple wireless LANs form an overall large network whose links interact and compete for airtime using the carrier-sense multiple access (CSMA) protocol. When a station hears its neighbors transmit, it will refrain from transmitting in order to avoid packet collisions.

Typically, the carrier-sense relationship between links can be "non-all-inclusive" in that each link only hears a subset of the links in the network and different links hear different subsets. In such a network, each link may experience different throughput since different links meet different situations when competing for the channel airtime. On the other hand, each link in the network interacts with each other, constituting an inseparable large-scale CSMA wireless network. Throughput analysis and computation of networks of this kind can be rather challenging and has attracted much attention from the community [1]–[4].

Ref. [3] presented an analytical model, Ideal CSMA Network (ICN), to study the behavior of saturated CSMA wireless networks. It was shown that the throughputs of links can be computed from the stationary probability distribution of the states of a continuous-time Markov chain. Although the ICN model has captured the main features of the CSMA protocol, an important simplifying assumption made in ICN is that each link is *saturated* in the network. In particular, it is assumed that the offered-load of each link is *infinite* so that each link always has a packet in its transmit buffer and competes for the channel. However, the traffic load in practical CSMA networks is *finite* and the transmit buffer of links can be empty from time to time [5]. For example, data traffic such as FTP download and web browser is typically bursty and real-time application such as voice over IP often operate at low rates. As far as we know, there has been little research on the "unsaturated non-all-inclusive" CSMA wireless networks. An outstanding issue in the literature is how to analyze the throughput performance of the CSMA protocol under finite offered-load traffics and this paper attempts to fill this gap.

In this paper, we propose an approximate approach for the throughput analysis of general CSMA wireless networks, in which links have finite offered-load and their transmit buffers may be empty at times. We use the "Equivalent Access Intensity" (EAI) to characterize the impacts of empty transmit buffers on the behavior of link channel competition and the interactions among links in the network. The access intensity of a link is the ratio of the mean packet transmission time (packet duration) to the mean backoff countdown time. When a link is unsaturated, since it will not perform channel competition when it has no packet waiting for transmission, the empty transmit buffer will incur "extra" waiting time for the link to compete for channel airtime usage. The basic idea of our approach is that this extra waiting time can be mapped to an equivalent "longer" backoff countdown time for each unsaturated link. By doing so we convert the unsaturated CSMA network to a saturated CSMA network in which the unsaturated links have "equivalent access intensities". Then, we can adopt the saturated ICN analysis for link throughput computation.

More specifically, we propose an iterative algorithm, "Compute-and-Compare", to identify which links are unsaturated under current offered-load and protocol settings. Note that this identification is non-trivial. Whether a link is saturated depends not only on its local offered load but also the throughputs it obtains in the network. It is known that each link in a CSMA network could affect each other and a link may be affected by another link far away [3]. Thus, we have to study the behavior of all links in overall network and then identify the unsaturated links. Second, for each unsaturated link we properly compute its equivalent access intensity by mapping the waiting time occurred by empty transmit buffer to the sum of its extra backoff time and the associated frozen time. This mapping constructs a saturated CSMA network in which the unsaturated links have equivalent access intensities. Finally, the link throughputs can be effectively computed using the ICN model under the constructed saturated CSMA network. Simulation bore out the accuracy of our EAI method. We believe the ability of link throughputs computation will serve as an important step toward the design and optimization of general CSMA wireless network with finite offered-load.

**Related Work**

Throughput analysis of CSMA wireless networks has attracted much attention in the literature and this is indeed a "hot" topic among researchers these years. Bianchi [6] proposed a two-dimensional model for IEEE 802.11 networks, assuming all links can sense each other. Following this model, much prior work studied the throughput, delay and fairness of such 802.11 networks. In particular, [7] extended Bianchi's model to the unsaturated case. Specially, an "idle" state was added into the Markov chain to capture the phenomena that the link has no packet waiting for transmission and quit channel competition. Ref. [8] extended Bianchi's model by incorporating the channel state and unsaturated links. Ref. [9] developed a buffer model to analyze a non-saturated IEEE 802.11 DCF networks and capture the influence of the transmit buffer size on the network performance. The authors of [10] derived the contention window sizes that maximize the WLAN system throughput under both saturated and non-saturated conditions. However, most prior work considered the "all-inclusive" case in which each link can sense each other and then has the same experience when it competes for the channel airtime. As can be seen in the technical report version of [3], it is extremely difficult to extend the method in [6] for the "non-all-inclusive" case.

Observing this, we proposed the analytical model ICN in [3] to study the interactions and competitions in a "non-all-inclusive" CSMA wireless network. Recent works that advanced the understanding and optimization of the "non-all-inclusive" CSMA networks include [11]–[13]. However, an assumption made in prior works is that the offered-load is infinite and then all links in the network are saturated. Noting that the saturation assumption is unlikely to be valid in real networks, this paper attempts to remove this assumption and to characterize the interactions and dependencies of links in the CSMA network in which certain links may quit channel competition due to no packet to transmit. We propose the "Equivalent Access Intensity" approach to capture the behavior of unsaturated links. More specifically, we develop the "Compute-and-Compare" algorithm to identify the unsaturated links and map the *unsaturated* network to a *saturated* network with the access intensities of the unsaturated links properly computed. Simulation results show that the algorithm has high accuracy under various offered-load and protocol settings.

The remainder of the paper is organized as follows. Section II introduces our system model and reviews its saturated equilibrium analysis. Section III presents the idea of the equivalent access intensity method, then shows the iterative algorithm to identify unsaturated links and compute the corresponding EAI. Experimental evaluations are performed in Section IV to examine the accuracy of our EAI method and finally section V concludes this paper.

## II. System Model and Saturated Throughput Analysis

In this section, we introduce our system model of a CSMA wireless network with general offered-load and briefly review the saturated throughput analysis of ICN [3]. The ICN model was used in several prior investigations [1], [2], [11][1].

### A. System Model

The carrier sensing relationship among links in a CSMA network is described by a contention graph $G = (V, E)$ [2]. Each link (a transmitter and receiver pair) is modeled as a vertex $i \in V$. Edges, on the other hand, model the carrier-sensing relationships among links. There is an edge $e \in E$ between two vertices if the transmitters of the two associated links can sense each other. We say two links are neighbors if there is an edge between them.

At any time, a link is in one of two possible states, active or idle. A link is active if there is a data transmission between its two end nodes. Thanks to carrier sensing, any two links that can hear each other will refrain from being active at the same time. A link sees the channel as idle if and only if none of its neighbors is active.

A link will not begin channel competition until it has a packet to transmit in its transmit buffer. Each link with a packet to transmit maintains a *backoff* timer, $C$, the initial value of which is a random variable with an *arbitrary* distribution $f(t_{cd})$. The timer value of the link decreases in a continuous manner with $dC/dt = -1$ as long as the link senses the channel as idle. If the channel is sensed busy (due to a neighbor transmitting), the countdown process is frozen and $dC/dt = 0$. When the channel becomes idle again, the countdown continues and $dC/dt = -1$ with $C$ initialized to the previous frozen value. When $C$ reaches 0, the link transmits a packet. The transmission duration is a random variable with arbitrary distribution $g(t_{tr})$. After the transmission, if there is no packet in its buffer to transmit, the link quits channel competition and waits for the arrival of the next packet. Otherwise, the link resets $C$ to a new random value according to the distribution $f(t_{cd})$, and the process repeats. We define the *access intensity* of a link as the ratio of its mean transmission duration to its mean backoff time: $\rho = E[t_{tr}]/E[t_{cd}]$.

We assume each link has an infinite transmit buffer. The packet arrival process of link $i$ is Poisson with rate $f_i$. The offered-load of overall network is $F = [f_1, f_2, \cdots, f_N]$, where $N$ is the number of links in the network. Let $s_i \in \{0, 1\}$ denote the state of link $i$, where $s_i = 1$ if link $i$ is active (transmitting) and $s_i = 0$ if link $i$ is idle (no packet to transmit, actively counting down or frozen). The overall **system state** of ICN is $s = s_1 s_2 ... s_N$. Note that $s_i$ and $s_j$ cannot both be 1 at the same time if links $i$ and $j$ are neighbors because (i) they can sense each other; and (ii) the probability of them counting down to zero and transmitting together is 0 under ICN (because the backoff time is a continuous random variable)[3].

---

[1]The correspondence between ICN and the IEEE 802.11 protocol [14] can be found in [3]

[2]Note that the pairwise carrier-sensing model is used in this paper. It is known that the cumulative interference model (or physical interference model) captures packet corruption more accurately as compared to the pairwise interference model. However, it is possible to implement pairwise carrier sensing (as modeled by the contention graph) that can prevent collisions under the cumulative interference model [15]. In this case, hidden-node collisions can be effectively eliminated.

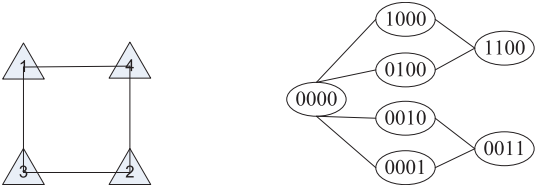

Fig. 1. (a) An example contention graph and (b) its state-transition diagram.

The collection of feasible states corresponds to the collection of independent sets of the contention graph. An independent set (IS) of a graph is a subset of vertices such that no edge joins any two of them [16].

As an example, Fig. 1(a) shows the contention graph of a network consisting of four links. In this network, links 1 and 2 can sense links 3 and 4 while links 3, 4 can hear links 1 and 2. Fig. 1(b) shows the associated state-transition diagram under the ICN model. To avoid clutters, we have merged the two directional transitions between two states into one line in Fig. 1(b). Each transition from left to right corresponds to the beginning of the transmission of one particular link, while the reverse transition corresponds to the ending of the transmission of that link. For example, the transition $1000 \rightarrow 1100$ is due to link 2's beginning to transmit; the reverse transition $1100 \rightarrow 1000$ is due to link 2's completing its transmission.

### B. Saturated Equilibrium Analysis

This part is a quick review of the saturated result in [3], and the reader is referred to [3] for details. We make two assumptions below:

- Assumption 1: The offered-load of each link is infinite so that each link always has a packet waiting for transmission in its buffer. In other words, all links in the network are saturated.
- Assumption 2: The distributions of the backoff countdown time and transmission time, i.e., $f(t_{cd})$ and $g(t_{tr})$ are exponential.

Given the above two assumptions, $s(t)$ is a *time-reversible* Markov process. For any pair of neighbor states in the continuous-time Markov chain, the transition from the left state to the right state occurs at rate $\lambda = 1/E[t_{cd}]$, and the transition from the right state to the left state occurs at rate $\mu = 1/E[t_{tr}]$.

Let $S$ denote the set of all feasible states, and $n_s$ be the number of transmitting links when the system is in state $s =$

[3]The readers are referred to Section II-A of [4] for a brief discussion of collision effects in CSMA networks

$s_1 s_2 ... s_N$. The stationary distribution of state $s$ can be shown to be:

$$P_s = \frac{\rho^{n_s}}{Z} \forall s \in S, \quad \text{where} \quad Z = \sum_{s \in S} \rho^{n_s} \quad (1)$$

The fraction of time during which link $i$ transmits is

$$th_i = \sum_{s:s_i=1} P_s, \quad (2)$$

which corresponds to the normalized throughput of link $i$.

Note that $Z$ in (1) is a weighted sum of independent sets of $G$. In statistical physics, $Z$ is referred to as the partition function and the computation of $Z$ is the crux of many problems, which is known to be NP-hard [16].

For the case where different links have difference access intensities, (1) can be generalized by replacing $\rho^{n_s}$ with the $\prod_{i:s_i=1 \text{ in } s} \rho_i$, where $\rho_i$ is the access intensity of link $i$.

Ref. [3] showed that (1) is in fact quite general and does not require Assumption 2. In particular, (1) is insensitive to the distribution of the transmission duration $g(t_{tr})$, and the distribution of the backoff duration $f(t_{cd})$, given the ratio of their mean $\rho = E[t_{tr}]/E[t_{cd}]$.

Assumption 1, however, cannot be removed from the ICN model. For example, recent work [11] makes (1) valid in their algorithm design by assuming that links transmit "dummy" packets when no packet in the buffer to transmit. The analysis and optimization of unsaturated CSMA networks is an outstanding issue in the literature. To analyze the CSMA networks in which links may be unsaturated, in Section III we propose the EAI method to characterize the impacts of empty transmit buffers on the behavior of link channel competition and the interactions among links in the network.

## III. Equivalent Access Intensity of Unsaturated links

In this section, we first elaborate the basic idea behind the EAI method in Part A. Part B provides the iterative algorithm, "Compute-and-Compare" to identify the unsaturated links under current offered-load and protocol settings. Meanwhile, the "Compute-and-Compare" algorithm yields the equivalent access intensities of the unsaturated links and the link throughputs of the overall network.

### A. Equivalent Access Intensity (EAI) Method

The basic idea of Equivalent Access Intensity method is as follows: in a saturated network [3], the state of a particular link $i$ alters between transmission, active countdown and frozen due to its neighbors' transmission. The access intensity of link $i$ is the ratio of its mean transmission time to its mean backoff time, $\rho_i = E[t_{tr}]/E[t_{cd}]$. When the link is unsaturated, at times it has to be idle and wait for the next packet's arrival at its transmit buffer. This waiting time could be mapped to an "extra" countdown backoff time (the mapping method will be introduced in Part B), leading to an equivalent access intensity of the link, $\tilde{\rho}_i = E[t_{tr}]/E[t_{cd} + \Delta t_{cd}]$, in which $\Delta t_{cd}$ is the extra countdown time mapped from the waiting time due to empty buffer. Then we could regard the link as saturated and it competes the channel airtime with the computed access intensity $\tilde{\rho}_i$.

Systematically, we consider two networks with the same network contention graph:

**Network 1:**

Network 1 is an unsaturated CSMA wireless network with offered load $F = [f_1, f_2, \cdots, f_N]$. The link access intensity is $P = [\rho_1, \rho_2, \cdots, \rho_N]$, which is determined by the packet duration and the backoff times the links adopts. Denote the throughputs of links in Network 1 by $TH = [th_1, th_2, \cdots, th_N]$.

**Network 2:**

We construct a saturated CSMA network with infinite offered load, Network 2. In this network, the link access intensity is $\widetilde{P} = [\widetilde{\rho}_1, \widetilde{\rho}_2, \cdots, \widetilde{\rho}_N]$, which is properly computed to make sure that the link thoughts equal to those of Network 1. That is, we want to find a set of link access intensities with which the link throughput of Network 2 is also $TH = [th_1, th_2, \cdots, th_N]$.

Theorem 1 below shows that the set of equivalent link access intensity $\tilde{P}$ exists in general. Assuming that we could find $\tilde{P}$ (An approximate computation will be introduced in Part B.), the construction of Network 2 enables us to compute the link throughputs of an unsaturated CSMA network using the saturated analysis, i.e., (1) and (2).

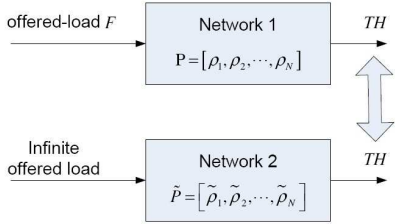

Fig. 2. The correspondence between Network 1 and Network 2.

*Theorem 1:* Suppose that the offered-load vector $F$ yields output-throughput vector $TH(P, F)$ under access-intensity vector P. Then, there exists another access-intensity vector $\tilde{P} = [\tilde{\rho}_i]$ under which the saturated-throughput vector is $TH(\tilde{P}) = TH(P, F)$.

*Proof:* The proof is relegated to the Appendix. ∎

Theorem 1 shows that the desired access-intensity vector $\tilde{P}$ does exists. However, we do not know how to compute it exactly. In Part B, we propose an approximate method to iteratively identify the subset of unsaturated links and compute their equivalent access intensities. Simulations in Section IV show this algorithm is of high accuracy under various simulation settings.

### B. Iterative Algorithm to Compute $\tilde{P} = [\tilde{\rho}_i]$

In this subsection, we present an iterative algorithm, "Compute-and-Compare", to identify the set of links that are unsaturated under current offered-load and protocol settings, compute the equivalent access intensities of these links and then obtain link throughputs of the overall network using saturated analysis.

Let us look at (1) and (2), a set of access intensity $P = [\rho_1, \rho_2, \cdots, \rho_N]$ corresponds to a set of $P_s$ and normalized link throughputs $TH = [th_1, th_2, \cdots, th_N]$. In the initialization procedure, we assume the network is saturated and compute the saturated throughput $TH^0 = \left[th_1^0, th_2^0, \cdots, th_N^0\right]$. Comparing $TH^0$ to $F$ [4], we can identify the subset of unsaturated links (if exist). We note that if a link is unsaturated, its output throughput must be equal to its offered-load; if a link is saturated, it will compete for the channel with its access intensity $\rho_i$ and spend no time in waiting for packet arrival. Hence, for the unsaturated link, we set its normalized throughput to its offered-load; for the saturated link, we set its access intensity to $\rho_i$. By combining (1) and (2), we can solve for a set of "equivalent" link access intensities for unsaturated links and a set of throughputs for saturated links. With the newly computed link throughputs, we re-identify the set of unsaturated links, repeat the "compute-and-compare" process until the set of unsaturated link over two successive iterations does not change. The detailed iterative algorithm is shown in Algorithm 1 below.

We next an example to illustrate the algorithm above: look at the CSMA network in Fig. 1(a). Let the link access intensity be $P = [\rho_0, \rho_0, \rho_0, \rho_0]$, where $\rho_0 = 5.3548$ is the link access intensity in a typical 802.11b networks and the corresponding saturated throughput $TH^0 = [0.4266, 0.4266, 0.4266, 0.4266]$. We next compute the output link throughputs under offered-load vector $F = [0.2, 0.4, 0.4266, 0.4266]$.

In the Initialization phase, since $F < TH^0$, the network is unsaturated. According to Step 10, $U^1 = \{1, 2\}$ and $S^1 = \{3, 4\}$. Then we invoke the EAI Computation procedure. In Step 17, we set $th^1 = f_1 = 0.2$, $th^2 = f_2 = 0.4$, $\widetilde{\rho}_3^1 = \widetilde{\rho}_4^1 = \rho_3 = \rho_0$ to find $\widetilde{\rho}_1^1, \widetilde{\rho}_2^1$ to achieve $th^1$ and $th^2$. By solving (1) and (2), we have $\widetilde{P}^1 = [0.8798, 14.6341, 5.3548, 5.3548]$. Since $\widetilde{\rho}_2^1 > \rho_2$, link 2 is actually saturated (for an unsaturated link, we have $\tilde{\rho}_i < \rho_i$). Hence, $S^1 = \{2, 3, 4\}$ and $U^1 = \{1\}$.

Repeat Step 17, we set $th^1 = f_1 = 0.2$ and we find $\widetilde{\rho}_1^1$ to meet it. By solving (1) and (2), we have $\widetilde{P}^2 = [1.7994, 5.3548, 5.3548, 5.3548]$ and $TH^1 = [0.2, 0.2622, 0.5952, 0.5952]$ under $\widetilde{P}^2$. Compared with $F$, we have $S^2 = \{2\}$ and $U^2 = \{1, 3, 4\}$. Repeat Step 17, we obtain $\widetilde{P}^2 = [0.7688, 5.3548, 2.7667, 2.7667]$ and $TH^2 = [0.2, 0.3877, 0.4266, 0.4266]$. Compared with $F$, $S^3 = S^2 = \{2\}$ and $U^3 = U^2 = \{1, 3, 4\}$. Thus, the program is terminated and the output throughput is $TH = [0.2, 0.3877, 0.4266, 0.4266]$. Our simulation shows that this computation is exact for this specific example. More simulation results will be presented in Section IV.

### C. Complexity of "Compute-and-Compare" Algorithm

The "Compute-and-Compare" Algorithm requires the ICN computation in each iteration. As stated in [3], the computation of ICN is an NP-complete problem. For modest-size CSMA networks, it can be realized by computer programs. For larger networks (e.g., networks of more than 100 links), the computation could be rather time-consuming. The algorithm proposed in this paper requires rounds of ICN computation and its computation could be out-of-control for general CSMA networks. Recently, [4] proposed to approximately compute link throughputs under the framework of belief propagation. As shown there, BP algorithms have good performance in terms of both speed and accuracy. It would be interesting

---
[4] We say $F \geq TH^0$ if $F, TH^0$ satisfy $\min_i(f_i - th_i) \geq 0$; $F < TH^0$ if $\max_i(f_i - th_i) \leq 0$ and $\exists i, th_i - f_i < 0$ are satisfied.

**Algorithm 1** "Compute-and-Compare" procedure for Throughput Computation of CSMA Wireless Network with Finite Offered-load

1: Consider a CSMA network with contention graph $G$ and access intensity $P = [\rho_1, \rho_2, \cdots, \rho_N]$. Suppose the offered load vector is given by $F = [f_1, f_2, \cdots, f_N]$. This algorithm computes the link throughputs $TH = [th_1, th_2, \cdots, th_N]$ of such a network.

2: By comparing the offered-load with the achievable saturated link throughputs, we iteratively identify which links are unsaturated, update their equivalent access intensities and recalculate the link throughputs under the new access intensity vector, until that the set of unsaturated links does not change over two successive iterations. Denote the equivalent link access-intensity vector and link throughput in the $n^{th}$ iteration by $\widetilde{P}^n = \left[\widetilde{\rho}_1^n, \widetilde{\rho}_2^n, \cdots, \widetilde{\rho}_N^n\right]$ and $TH^n = \left[th_1^n, th_2^n, \cdots, th_N^n\right]$, respectively. Furthermore, we denote the set of saturated links and unsaturated links in the $n^{th}$ iteration by $S^n$ and $U^n$, respectively.

3: **procedure** INITIALIZATION
4:     We set $\widetilde{P}^0 = P = [\rho_1, \rho_2, \cdots, \rho_N]$, $S^0 = V$ and $U^0 = \emptyset$.
5:     Using $\widetilde{P}^0$, calculate the link throughput with (1) and (2). Denote the computed throughput by $TH^0 = \left[th_1^0, th_2^0, \cdots, th_N^0\right]$.
6:     **if** $F \geq TH^0$ **then**
7:         The network is saturated and $TH^0$ is the output link throughputs, i.e., $TH = TH^0$.
8:         Terminate the program.
9:     **else**
10:        The network is unsaturated; Let $S^1 = \left\{i | f_i \geq th_i^0\right\}$ and $U^1 = \left\{i | f_i < th_i^0\right\}$.
11:        $n = 1$.
12:        Invoke procedure EAI COMPUTATION.
13:     **end if**
14: **end procedure**

15: **procedure** EAI COMPUTATION
16:     **while** $S^n \neq S^{n-1}$ (i.e., $U^n \neq U^{n-1}$) **do**
17:         Set the target throughputs of the links in $U^n$ to $f_i$ and fix the access intensities of links in $S^n$ to $P_{S^n} = [\rho_1, \rho_2, \cdots, \rho_{|S^n|}]$, combining (1) and (2) we solve for the access intensities of links in $U^n$ to meet their output throughputs and update $\widetilde{P}^n = \left[\widetilde{\rho}_1^n, \widetilde{\rho}_2^n, \cdots, \widetilde{\rho}_N^n\right]$ accordingly. Note that for each link $i \in S^n$, we have $\widetilde{\rho}_i^n = \rho_i$.
18:         If $\widetilde{P}^n \leq P$, go to Step 19; otherwise, denote the set the links with access intensity larger than that in P by $\Delta U$. Note that the access intensity of a link in unsaturated situation cannot be larger than that in saturated case. Hence, the links in $U^n$ whose access intensities are larger than their access intensities in P must be saturated. Thus, $S^n \leftarrow S^n + \Delta U$ and $U^n \leftarrow U^n - \Delta U$, go to Step 17.
19:         Using $\widetilde{P}^n = \left[\widetilde{\rho}_1^n, \widetilde{\rho}_2^n, \cdots, \widetilde{\rho}_N^n\right]$, compute the link throughputs $TH^n = \left[th_1^n, th_2^n, \cdots, th_N^n\right]$.
20:         Let $S^{n+1} = \left\{i | f_i > th_i^n\right\}$ and $U^{n+1} = \left\{i | f_i \leq th_i^n\right\}$.
21:         $n = n + 1$.
22:     **end while**
23:     The output link throughput $TH = TH^{n-1}$.
24: **end procedure**

to explore how to tailor the BP algorithm to tackle the high complexity issues of the "Compute-and-Compare" algorithm proposed in this paper. We have obtained preliminary results and will be presented somewhere else to limit the scope of this paper.

## IV. EXPERIMENTAL EVALUATION

We conduct simulations to examine the accuracy of the EAI computation. We implement Algorithm 1 and ICN-simulator with finite offered-load using MATLAB programs. Table I shows the results of EAI computation for various network topologies and offered-load vectors. Typical link access intensity $\rho_0 = 5.3548$ was used in the simulations.

As can be seen in Table I, the EAI computation is of high accuracy. EAI is also borne out by simulations of some other networks of similar size with various offered-load settings that are not shown here.

For larger networks, we run three sets of 10 randomly generated 20-link networks. The mean degree of links (i.e., the number of neighbors per link in the contention graph) is 2, 3 and 4, respectively. To generate an unsaturated offered-load vector, we run the saturated ICN simulator using $\rho_0$ to get the saturated link throughputs $TH^0$. Label the links by $1, 2, \cdots, N$, for odd links, we set $f_i = th_i$ while for even links, we set $f_i = \max(th_i - 0.1, 0)$. By doing so, we construct an unsaturated offered-load vector and set it to be the offered-load of ICN simulation. Then we perform the EAI computation and compare its results with the throughputs obtained from simulation. For each link, we calculate the error of the throughput obtained by the ICN-simulator relative to the EAI computation. Table II shows the mean link throughput errors of EAI for different link-degrees. As can be seen there, the error of EAI averaged the 10 runs is within 0.1%, which is quite amazing.

## V. CONCLUSION

This paper has proposed an approximate method (EAI) to analyze the link throughputs of a CSMA network with finite offered-load. To the best of our knowledge, this is the first attempt to analyze the link throughputs of an unsaturated "non-all-inclusive" CSMA wireless network. The impacts of empty transmit buffer on the behavior of the unsaturated link and interaction among links have been well captured. An iterative algorithm was proposed to identify the subset of unsaturated links and compute link throughputs. Simulations bore out the accuracy of the proposed EAI method.

We believe the EAI method enriches our understanding on unsaturated CSMA networks. For example, based on this model, we could design algorithms to optimize CSMA network by offered-load control instead of adjusting link access-intensities (e.g., the two networks in Fig.2 have the same link throughputs). It is worthwhile to note that offered-load control is superior to link access intensity adjustment as far as engineering implementation is considered. In practical networks, the packet arrival rate of the MAC-layer transmit buffer (i.e., the offered-load) can be controlled by the application layer through software programs. In contrast, the link access intensity is a parameter of the CSMA protocol that has been fixed and cannot be modified by the driver in most of business 802.11 cards.

TABLE I
CONTENTION GRAPHS OF VARIOUS NETWORK TOPOLOGIES AND OFFERED-LOADS, THE CORRESPONDING EAI-COMPUTED RESULTS AND SIMULATION RESULTS.

| Network Topology | Offered-load | EAI | Simulation |
|---|---|---|---|
| 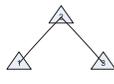 Topology 1 | [0.4266, 0.4266, 0.4266] | [0.4266, 0.3653, 0.4266] | [0.4258, 0.3627, 0.4286] |
| | [0.2000, 0.2000, 1.0000] | [0.2000, 0.2000, 0.6741] | [0.1992, 0.2000, 0.6739] |
| 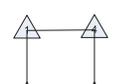 Topology 2 | [0.2000, 0.4266, 0.4266, 0.4266] | [0.2000, 0.3877, 0.4266, 0.4266] | [0.1996, 0.3880, 0.4278, 0.4263] |
| | [0.2000, 0.4000, 0.4266, 0.4266] | [0.2000, 0.3877, 0.4266, 0.4266] | [0.1999, 0.3779, 0.4270, 0.4262] |
| 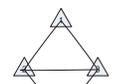 Topology 3 | [0.5500, 0.1048, 0.1048, 0.6660] | [0.5500, 0.1048, 0.1048, 0.6660] | [0.5510, 0.1057, 0.1052, 0.6638] |
| | [0.4000, 0.3000, 0.3000, 0.4000] | [0.4000, 0.224, 0.224, 0.4000] | [0.4025, 0.2133, 0.2135, 0.3992] |

TABLE II
MEAN LINK THROUGHPUT ERRORS OF EAI, FOR NETWORKS WITH DIFFERENT MEAN LINK DEGREES.

| Mean Link Degree | 2 | 3 | 4 |
|---|---|---|---|
| Mean Throughput Errors | 0.048% | 0.064% | 0.091% |

APPENDIX

In this appendix, we prove Theorem 1.

*Proof:* 1) To show that $TH(\mathrm{P}, F)$ satisfies $\exists P_s, P_s > 0$ and $\sum_s P_s = 1$, such that $TH(\mathrm{P}, F) = \sum_s P_s s$, where $s$ is a feasible state and $P_s$ is the airtime dedicated to state $s$.

Suppose $F > 0$ (If $f_i = 0$, this means the link has no packet to transmit and will be always idle. Hence, we remove the link and its associated edges from the network contention graph.). Thus, in this network, with probability 1 link $i$ will transit to the actively countdown or frozen state, starting from the state of waiting for the next packet to arrive at its buffer. That is, the system process will not be stuck by the state in which $s_i = 0$.

Let us look at the states of the network under offered-load $F$. At any time it must be in one and only one specific state $s$ and $\sum_s P_s = 1$. Next we show that in equilibrium each state has non-zero probability by shown that $\forall s, s' \in \mathcal{S}$, $s$ and $s'$ communicate with each other. To see this, we transform the state transition diagram of the CSMA network to a directed graph $G'$, in which the vertices represent the system states and the arcs represent the transition edges in the state transition diagram, respectively. Now it is sufficient to show that there exists a path between $s$ and $s'$ on $G'$. Let $m = SP(s, s')$ denote the length of the shortest path for the process to reach state $s'$ starting from state $s$. Then construct a path composed of a sequence of states $s^1, s^2, \cdots, s^{m-1}$, such that $SP(s', s^1) = m - 1$, $SP(s', s^2) = m - 2$, $\cdots$, $SP(s', s^{m-1}) = 1$. That is, starting from $s$, the process approaches state $s'$ by changing one element corresponding to state $s$ upon each transition. Thus, $TH(\mathrm{P}, F)$ satisfies $\exists P_s, P_s > 0$ and $\sum_s P_s = 1$, such that $TH(\mathrm{P}, F) = \sum_s P_s s$.

2) Invoking Proposition 2 of [11], any strictly feasible offered-load $F$ is attainable by finite link access intensities in the saturated network. Thus, there exists a link access-intensity vector $\tilde{\mathrm{P}}$, that yields saturated throughput $TH(\mathrm{P}, F)$. ∎